\documentclass[pdflatex]{sn-jnl}

\usepackage{graphicx}%
\usepackage{multirow}%
\usepackage{amsmath,amssymb,amsfonts}%
\usepackage{amsthm}%
\usepackage{mathrsfs}%
\usepackage[title]{appendix}%
\usepackage{xcolor}%
\usepackage{textcomp}%
\usepackage{manyfoot}%
\usepackage{booktabs}%
\usepackage{algorithm}%
\usepackage{algorithmicx}%
\usepackage{algpseudocode}%
\usepackage{listings}%
\begin{document}

\title[Theoretical and computational models for Saturn's\\ co-orbiting moons, Janus and Epimetheus]{Theoretical and computational models for Saturn's\\ co-orbiting moons, Janus and Epimetheus}

\author*[1]{\fnm{Sean} \sur{O'Neill}}\email{oneillsm@plu.edu}

\author[1]{\fnm{Katrina} \sur{Hay}} 

\author[2]{\fnm{Justin} \sur{deMattos}}

\affil*[1]{\orgdiv{Physics Department}, \orgname{Pacific Lutheran University}, \orgaddress{\city{Tacoma}, \state{Washington}, \country{USA}}}

\affil[2]{\orgname{Northrop Grumman}, \orgaddress{\city{Redondo Beach}, \state{California}, \country{USA}}}

\abstract{Two moons of Saturn, Janus and Epimetheus, are in co-orbital motion, exchanging orbits approximately every four Earth years as the inner moon approaches the outer moon and they gravitationally interact. The orbital radii of these moons differ by only 50 km (less than the moons' mean physical radii), and it is this slight difference in their orbits that enables their periodic exchanges.  Numerical $n$-body simulations can accurately model these exchanges using only Newtonian physics acting upon three objects: Saturn, Janus, and Epimetheus.  Here we present analytical approaches and solutions, and corresponding computer simulations, designed to explore the effects of the initial orbital radius difference on otherwise similar co-orbital systems.  Comparison with our simulation results illustrates that our analytic expressions provide very accurate predictions for the moon separations at closest approach and simulated post-exchange orbital radii.  Our analytic estimates of the exchange period also match the simulated value for Janus and Epimetheus to within a few percent, although systems with smaller differences in their orbital radii are less well-modeled by our simple approach, suggesting that either full simulations or more sophisticated analytic approaches would be required to estimate exchange periods in those cases.}


\maketitle
We gratefully acknowledge Pacific Lutheran University for support on this project and CoCalc for the use of their computational resources.

\newpage


\section{Introduction}
\label{Introduction} 

In 1966, astronomers made telescope observations of two objects beyond the outer rings of Saturn, subsequently confirmed to be the two separate moons now known as Janus and Epimetheus (Dermott and Murray 1981). The orbits of the two moons were so similar to one another that it took over 10 years to establish definitively that they were not the same object.  Observations by Voyager 1 (Aksnes 1985), and later the Hubble Space Telescope and Cassini spacecraft (Kolhase and Peterson 1997; Spitale et al. 2006; Jacobson et al. 2008; Cooper et al. 2015), have given us valuable glimpses into the unusual orbits of these moons.
In detail, the average difference in the orbital radii of the two moons is approximately 50 km, which is less than the mean physical radius of either Janus (89.5 km) or Epimetheus (58.1 km) (Thomas 2010). The moons' orbits are so similar, in fact, that their orbital periods differ by less than an Earth minute per orbit, with each orbit taking approximately 17 Earth hours (Jacobson et al. 2008).  

How can Janus and Epimetheus occupy orbits in the same plane without colliding, despite having an orbital radius difference that is smaller than the moons themselves?  The answer is that they exhibit the phenomenon known as co-orbital motion (Dermott and Murray 1981; Yoder et al. 1983). Approximately every four Earth years, the moons exchange orbits, taking turns as the inner moon (closer to Saturn) and the outer moon (further from Saturn). This occurs due to their mutual gravitational interaction. According to Kepler's Laws, the inner moon moves faster tangentially in its orbit. As the inner moon approaches the outer moon, it pulls the outer moon down toward its orbit while the outer moon simultaneously pulls the inner moon up toward its orbit. What had been the outer moon remains in the lead, but has now become the faster-moving inner moon. Four years later, the exchange occurs again with the moons' roles reversed. 

\begin{figure}[ht!]
\centering
\includegraphics[width=12cm]{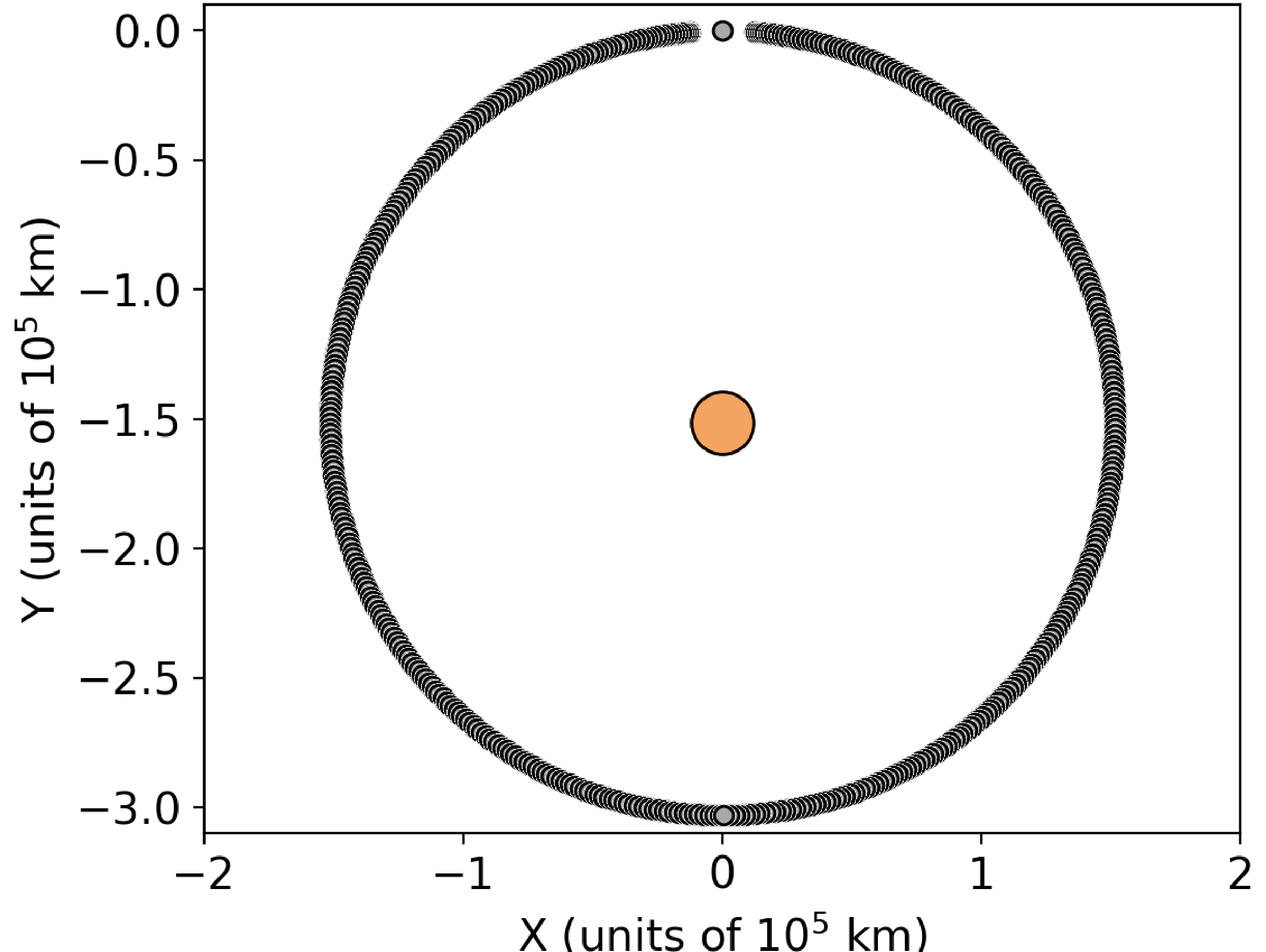}
\caption{An illustration of a horseshoe orbit.  In a non-inertial reference frame in which a vertical line connects Saturn (larger central circle) and one of the two moons (smaller upper circle), the orbit of the second moon traces out a ``horseshoe'' shape, shown here as a overlapping grey path.}
\label{fig_horseshoe}
\end{figure}

Notably, the inner moon never actually catches up to or passes the outer moon. Instead, the orbit exchange takes place before the two moons collide.  This particular type of co-orbital motion is sometimes called a ``horseshoe orbit," owing to the shape made by one of the co-orbiting moons, while observing the system from the rest frame of the other moon (see Figure \ref{fig_horseshoe}).  Janus and Epimetheus are the only solar system objects of comparable mass currently known to execute a full horseshoe orbit (Dermott and Murray 1981; Treffenst\"adt et al. 2015); El Moutamid et al. 2016). According to NASA's spacecraft observations, Janus and Epimetheus come within 15,000 km of one another (Llibre and Oll\'e 2001). In our subsequent discussion, we will refer to this distance as the ``closest approach."

It is known that horseshoe orbits can be stable (Murray and Dermott 1999; Yoder et al. 1983, Llibre and Oll\'e 2001; \'Cuk et al. 2012). Neiderman et.al. (2020) provide a proof of existence of orbit trajectories similar to Janus and Epimetheus to investigate the long term stability of such a system.  Vanderbei (2005) argues that the moons' horseshoeing behavior makes collisions uncommon, leading to a Saturn ring system that is more stable than previously believed because of the presence of co-orbiting moons. Theories for the formation of this co-orbiting system through collisions are described in the literature (Treffenst\"adt et al. 2015; Yoder et al. 1983). Alternatively, Vanderbei describes a process of horseshoe orbit formation through small perturbations that over time moved moons that once shared an orbit to slightly differing radii, setting the co-orbiting behavior into motion.

Orbital dynamics have been used to predict a range of distance and mass values that allow co-orbital behavior (Dermott and Murray 1981; Yoder et al. 1983 Cors and Hall 2003; Llibre and Oll\'e 2001; Neiderman et al. 2020). Cors and Hall (2003) studied constraints on mass and orbit shape based on data gathered from NASA mission observations. They provide a rigorous theoretical framework for the existence of co-orbital behavior and give a minimum angular separation for the horseshoe orbits based on satellite parameters, such as mass and orbital radii. 

In this work, we take a two-pronged approach to modeling the co-orbital motion of Janus and Epimetheus.  In Section \ref{analytical}, we describe a series of analytical approaches for estimating the exchange period, post-exchange orbital radii, and closest approach separation of two interacting moons co-orbiting a more massive planet.  In Section \ref{numerical}, we employ a series of $n$-body simulations of such systems designed to test our analytical models and to provide further insight into the phenomenon over a range of initial differences in orbital radius.  We discuss our results in Section \ref{results}, comparing the predictions of our analytic models to our simulations.  We present our conclusions and suggest future extensions of our work in Section \ref{conclusions}.

\section{Analytical Solutions} 
\label{analytical}
First, we discuss theoretical expectations for the exchange period, post-exchange orbital radii, and closest approach separation for a set of co-orbital moons.  We will show that these analytic solutions are a time-saving shortcut for deriving particular orbital parameters that would otherwise require lengthy $n$-body simulations.  Although the following exercises were performed with Janus and Epimetheus in mind, the results are presented in forms that generalize to any system of two co-orbiting moons in low-eccentricity orbits around a much more massive planet.  Graphs of our analytic results and comparisons with our numerical simulations of co-orbital motion appear in Section \ref{results}.

\subsection{Exchange Period}
\label{analytical_period}
 To construct an estimate of the co-orbital exchange period, we model the two moons as test particles that do not gravitationally interact with one another, using the time required for the inner moon to ``lap'' the outer moon as a proxy for the exchange period.  This is an overly simplistic assumption - clearly, the mutual gravitational interaction of the two moons is required for the exchange to take place - but the moons are far enough apart for a sufficiently large fraction of their orbits that this approach results in a reasonably accurate exchange period estimate for the orbits of Janus and Epimetheus.

In detail, Newton's form of Kepler's Third Law of Planetary Motion tells us that a moon $n$ orbiting a planet of mass $M_p$ will take a time $T_n$ to complete one orbit according to
\begin{equation}
T_n = \frac{2\pi}{\sqrt{GM_p}}r_n^{3/2}~,
\end {equation}
where $G$ is the gravitational constant, $M_p$ is the mass of the planet, and $r_n$ is the orbital radius of moon.  For two moons with different orbital radii, the positive difference in their orbital periods is $\Delta T_{\rm orb} = T_o - T_i$, where $T_o$ is the period of the outer moon and $T_i$ is the period of the inner moon.  $\Delta T_{\rm orb}$ represents the period difference per orbit, so the number of orbits required for the inner moon to lap the outer moon is simply $N_{\rm orb} = T_o/\Delta T_{\rm orb}$.  We can combine these expressions to calculate the time required for the inner moon to lap the outer moon:
\begin{equation}
  \label{eq_exchange_period}
  T_{\rm lap} = N_{\rm orb} T_i = \frac{T_oT_i}{T_o - T_i}=\frac{2\pi}{\sqrt{GM_p}} \frac{(r_or_i)^{3/2}}{(r_o^{3/2} - r_i^{3/2})}~.
  \end{equation}

\subsection{Post-Exchange Orbital Radii}
\label{analytical_radii}
Because Janus and Epimetheus have different masses, their exchanges are not perfectly symmetric, meaning that the pre-exchange orbital radius of the old inner moon is not identical to the post-exchange orbital radius of the new inner moon (and similarly for the outer moon).  To calculate expected values for the post-exchange orbital radii, we employ the conservation of mechanical energy and angular momentum as follows.

\begin{figure}[ht!]
\centering
\includegraphics[width=12cm]{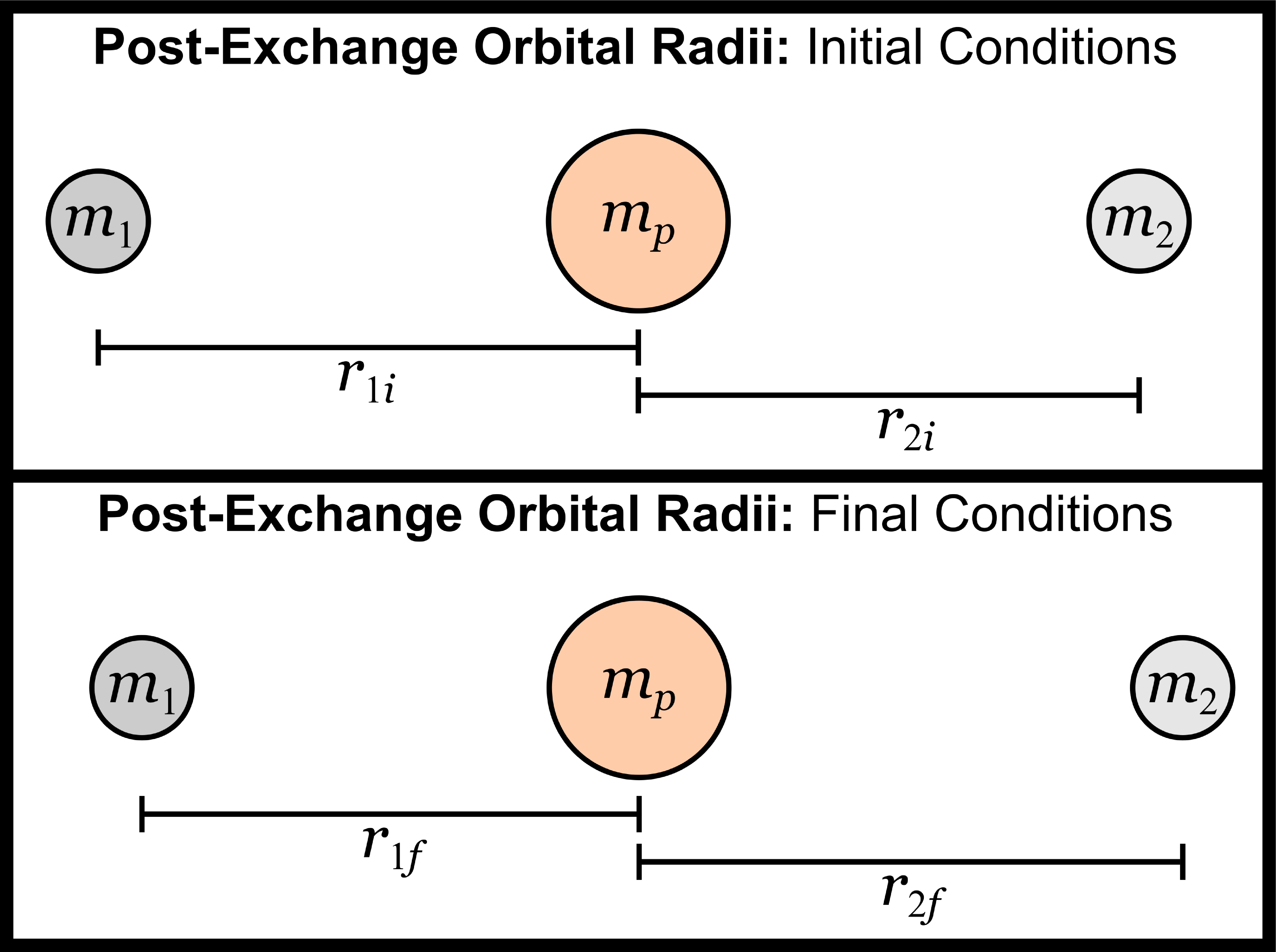}
\caption{Sample initial and final conditions for the post-exchange orbital radii calculation.  In this particular case, Moon 1 goes from being the outer moon (\textit{upper panel}) to being the inner moon (\textit{lower panel}), but the moons are on opposite sides of Saturn in both sets of conditions.  (Distances not to scale.)}
\label{radii_diagram}
\end{figure}

As shown in Figure \ref{radii_diagram},
the initial conditions (subscripted ``$i$'') correspond to the two moons being located on opposite sides of Saturn \textit{before} an exchange and the final conditions (subscripted ``$f$'') correspond to the two moons being located on opposite sides of Saturn \textit{after} an exchange.  In this configuration, both the gravitational potential energy between the two moons and Saturn's orbital kinetic energy around the system's center of mass are several orders of magnitude smaller than the other energy terms, allowing us to restrict our analysis to only the kinetic energies of the two moons and their gravitational potential energies with respect to Saturn.  The conservation of mechanical energy ($E_i=E_f$) therefore becomes

\begin{equation}
  \frac{1}{2}m_1v_{1i}^2 + \frac{1}{2}m_2v_{2i}^2 - \frac{G M_p m_1}{r_{1i}}-\frac{G M_p m_2}{r_{2i}} = \frac{1}{2}m_1v_{1f}^2 + \frac{1}{2}m_2v_{2f}^2 - \frac{G M_p m_1}{r_{1f}}-\frac{G M_p m_2}{r_{2f}}~,
  \end{equation}
where $v_{1i}$ corresponds to the initial speed of moon 1, etc.  Because Janus and Epimetheus feature orbits of very low eccentricity (Yoder et al. 1983), we can write their speeds using the form for a circular orbit: ($v_n = \sqrt{GM_p/r_n}$) and cancel terms to find that
\begin{equation}
  \frac{m_1}{r_{1i}}+\frac{m_2}{r_{2i}} = \frac{m_1}{r_{1f}}+\frac{m_2}{r_{2f}}~.
\end{equation}
We can simplify this expression by introducing a series of dimensionless variables
\begin{equation}
\label{eq_ratio_def1}
  \mu_2 \equiv \frac{m_2}{m_1}~~~,~~~\rho_{1f} \equiv \frac{r_{1f}}{r_{1i}}~~~,~~~\rho_{2i} \equiv \frac{r_{2i}}{r_{1i}}~~~,~~~\rho_{2f} \equiv \frac{r_{sf}}{r_{1i}}~,
\end{equation}
where each radius has been normalized by $r_{1i}$ to make this approach more generalizable to other systems.  The conservation of energy expression then becomes
\begin{equation}
  1 + \frac{\mu_2}{\rho_{2i}} = \frac{1}{\rho_{1f}}+\frac{\mu_2}{\rho_{2f}}~.
  \end{equation}
We apply a similar approach to angular momentum conservation ($L_i = L_f$), ignoring the orbital angular momentum of Saturn and assuming circular moon orbits, to find that
\begin{equation}
\label{eq_ang_mom1}
  1 + \mu_2\sqrt{\rho_{2i}} = \sqrt{\rho_{1f}} + \mu_2\sqrt{\rho_{2f}}~.
  \end{equation}
These two expressions can be combined into an equation of the form
\begin{equation}
  A\rho_{2f}^2 + B\rho_{2f}^{3/2} + C\rho_{2f}+D\rho_{2f}^{1/2}+E = 0 ~,
\end{equation}
where the coefficients are given by
\begin{equation}
\begin{aligned}
    A &= \mu_2\left(1 + \frac{\mu_2}{\rho_{2i}}\right) ~,\\
    B &=-2\left(1 + \frac{\mu_2}{\rho_{2i}}\right)\left(1+\mu_2\sqrt{\rho_{2i}}\right) ~,\\
    C &= \mu_2 \rho_{2i} + 2\sqrt{\rho_{2i}} + \frac{2\mu_2}{\sqrt{\rho_{2i}}}+ \frac{1}{\rho_{2i}}~,\\
    D &= 2\mu_2\left(1+\mu_2\sqrt{\rho_{2i}}\right) ~,~{\rm and}\\
    E  &= -\left(1+\mu_2\sqrt{\rho_{2i}}\right)^2 ~.
  \end{aligned}
  \end{equation}

The solutions to this equation are well-fit by a closed-form linear relationship between the dimensionless final and initial orbital radii, namely
\begin{equation}
    \label{eq_post_exchange_radius1}
    \rho_{1f} = (1-k)\rho_{2i} + k~~{\rm and}
  \end{equation}
  \begin{equation}
        \label{eq_post_exchange_radius2}
        \rho_{2f} = -k\rho_{2i} + (1+k) ~,
      \end{equation}
where $k$ is defined in terms of the moon mass ratio as
\begin{equation}
\label{eq_k_def}
  k=\frac{1-\mu_2}{1+\mu_2}~.\end{equation}
In summary, the closed-form Equations \ref{eq_post_exchange_radius1} and \ref{eq_post_exchange_radius2} can be used directly to calculate the post-exchange orbital radii, given the initial orbital radii and moon masses (as used in Equations \ref{eq_ratio_def1} and \ref{eq_k_def}).

\subsection{Closest Approach}
\label{analytical_closest_approach}
To calculate the co-orbital moon separation at closest approach, we again employ the conservation of mechanical energy and angular momentum.  Because their separation at closest approach involves the mutual gravitational potential between the two moons, however, we must now include this energy term explicitly in the final conditions as
\begin{equation}U_{12f} = -\frac{Gm_1m_2}{d^2}~,\end{equation}
where $d$ is the minimum separation between the two moons.  Additionally, because Saturn's orbital kinetic energy is coincidentally the same order of magnitude as $U_{12f}$, we must also account for that energy in the initial and final conditions using terms of the form
\begin{equation}K_{p} = \frac{1}{2}M_p v_{p}^2 ~,\end{equation}
where $v_p$ is the velocity of Saturn.
The kinetic energy terms for Saturn can be cast in terms of the moon orbit parameters by applying the conservation of linear momentum, which ensures that $p=m_1v_{1} + m_2v_{2}+M_pv_{p}=0$ for both initial and final conditions in our chosen reference frame.
\begin{figure}[ht!]
\centering
\includegraphics[width=12cm]{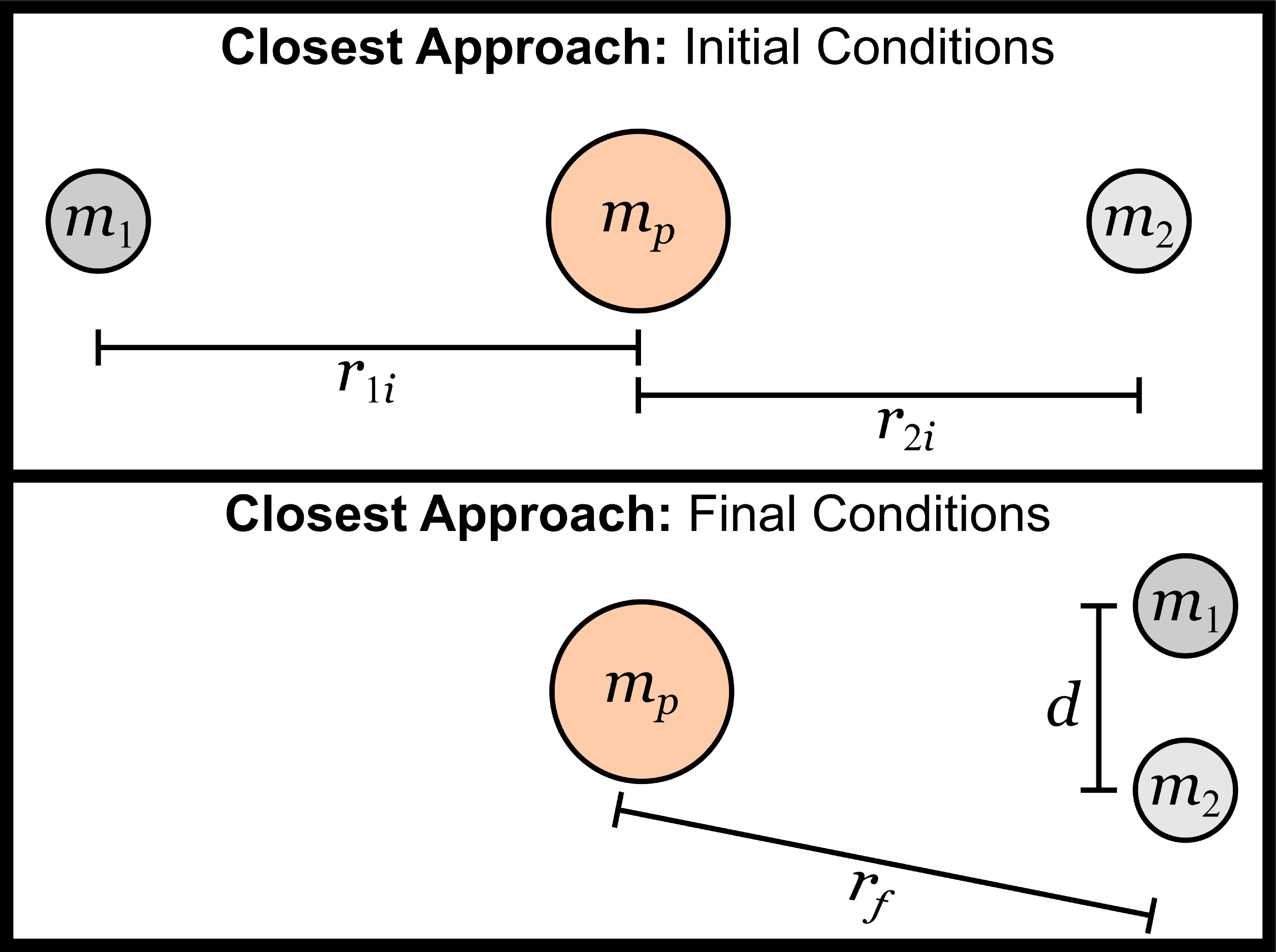}
\caption{Sample initial and final conditions for calculating the closest approach between the two moons.  The initial conditions assume that the moons are at opposite sides of Saturn, but the final conditions are chosen so that the moons are as close as possible to one another, separated by a distance $d$ and orbiting Saturn with identical orbital radii $r_f$.  (Distances not to scale.)}
\label{closest_approach_diagram}
\end{figure}

The final new assumption in this calculation is that the two moons have the same orbital speed at the point of closest approach, which is sensible since the moment of closest approach instantaneously represents a turnaround point for one of the moons in the other moon's rest frame.  For circular orbits, this means that the two moons must also have the same orbital radius at the point of closest approach, which we represent with a single variable $r_f$.  Combining all of these constraints and applying the initial and final conditions illustrated in Figure \ref{closest_approach_diagram} generates Equation \ref{eq_closest_approach1}, a more involved equation for the conservation of energy.  As we will discuss, this equation can be solved in combination with Equation \ref{eq_closest_approach2} quickly and easily via root-finding.
\[
  \frac{1}{\mu_p} - 1 + \left(\frac{\mu_2}{\rho_{2i}}\right)\left(\frac{\mu_2}{\mu_p}-1\right) - 2\left(\frac{\mu_2}{\mu_p}\right)\sqrt{\frac{1}{\rho_{2i}}} - 2\left(\frac{\mu_2}{\mu_p}\right)\left(\frac{1}{\rho_{12i}}\right)\]
\begin{equation}
\label{eq_closest_approach1}
  = \frac{(1+\mu_2)^2}{\mu_p}\left(\frac{1}{\rho_f}\right)-\frac{\mu_2}{\mu_p}\left(\frac{1}{\rho_f}\right)\left(\frac{\rho_{12f}}{\rho_f}\right)^2 - (1+\mu_2)\left(\frac{1}{\rho_f}\right)-2 \left(\frac{\mu_2}{\mu_p}\right)\left(\frac{1}{\rho_{12f}}\right) ~,\end{equation}
where we introduce the additional dimensionless variables
\begin{equation}\mu_p \equiv \frac{m_p}{m_1}~~~,~~~\rho_{f} \equiv \frac{r_{f}}{r_{1i}}~~~,~~~\rho_{12f} \equiv \frac{r_{12f}}{r_{1i}}~.\end{equation}
The equation representing angular momentum conservation is only slightly modified from Equation \ref{eq_ang_mom1}, becoming
\begin{equation}
\label{eq_closest_approach2}
  1 + \mu_2\sqrt{\rho_{2i}} = \sqrt{\rho_{f}}(1 + \mu_2) ~.\end{equation}
Unlike the situation for the post-exchange orbital radii, Equations \ref{eq_closest_approach1} and \ref{eq_closest_approach2} are not amenable to a linear solution.  Instead, these two equations must be solved simultaneously using common root-finding techniques in which we step through values of the post-exchange orbital radii (given initial orbital radii and moon masses) until a condition of equality is satisfied for both equations.  Typically, this takes a miniscule amount of computational time with a simple program, as compared to the significantly more computationally expensive full $n$-body simulations we present in Section \ref{numerical}.

\begin{figure}[ht!]
\centering
\includegraphics[width=12cm]{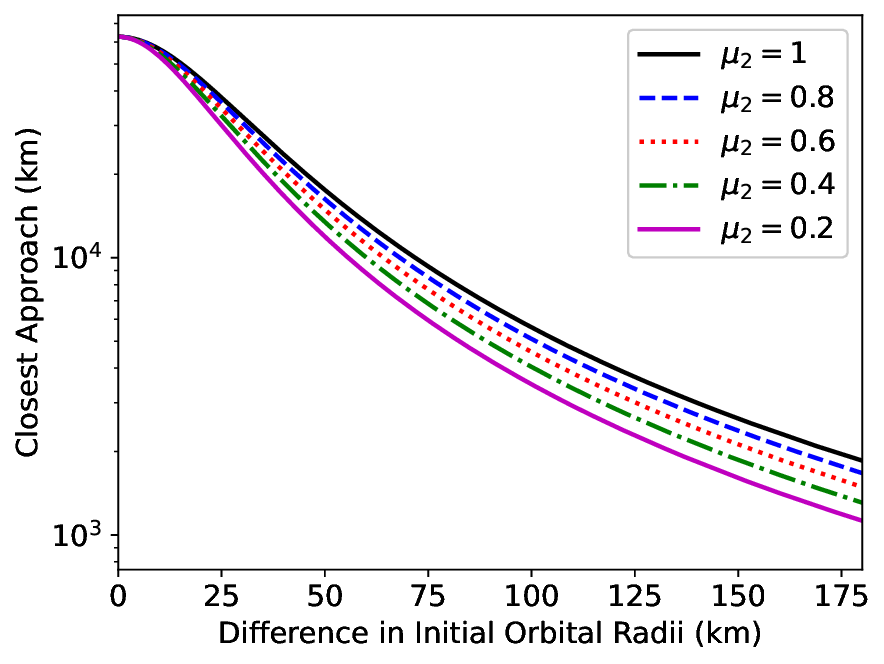}
\caption{Sample solutions to Equations \ref{eq_closest_approach1} and \ref{eq_closest_approach2} for a range of mass ratios $\mu_2$.}
\label{fig_root_curves}
\end{figure}

Whereas the previous solutions were presented in closed form, we present sample results from root-finding of Equations \ref{eq_closest_approach1} and \ref{eq_closest_approach2} in Figure \ref{fig_root_curves}.
These solutions are further compared to our simulations in Section \ref{results} for the specific case of Janus and Epimetheus.

\section{Numerical modeling of co-orbital motion} 
\label{numerical}

Direct numerical modeling of co-orbital moon behaviors facilitates testing of the analytical solutions presented in the previous section while also providing full orbital trajectories that are not easily derived from theoretical treatments of the three-body problem.  We therefore employed an $n$-body integrator that solves for the trajectories of pointlike objects under the influence of Newtonian gravity and mechanics (consult Aarseth 2003 or Dehnen and Read 2011 for general and technical discussions of $n$-body codes).  Briefly, the code calculates the net gravitational force on an object $i$ according to
\begin{equation}\vec{F}_{i,{\rm net}} = \sum_{j=1}^N -\frac{Gm_i m_j}{r_{ji}^2}\hat{r}_{ji}~,\end{equation}
where $G$ is the gravitational constant, $m_i$ is the mass of object $i$, $m_j$ is the mass of object $j$ located a distance $r_{ji}$ from object $i$, and the unit vector $\hat{r}_{ji}$ points from object $j$ to object $i$.  The acceleration of object $i$ is then calculated from $\vec{F}_{i,{\rm net}} = m_i \vec{a}_i$, which enables estimation of the subsequent velocity and position.  Specifically, this integrator employs the second-order leapfrog method (Aarseth 2003; Dehnen and read 2011), which uses an intermediate value of the velocity to advance the position by a timestep $\Delta t$.  This symplectic approach has the benefit of preventing the system energy from drifting over time, which is useful for orbits that deviate only slightly from perfectly periodic motion.  This particular $n$-body code also restricts its orbit calculations to two spatial dimensions, which is appropriate for Janus, Epimetheus, and Saturn, all of which move in approximately the same plane.

We conducted a set of 18 simulations of moons orbiting Saturn with a mass ratio equal to that of Janus and Epimetheus.  The simulations were identical to one another except for the moons' initial orbital radii, which differed from one another by amounts ranging from 10 - 175 km (including the actual value of approximately 50 km).  In each case, the two moons were initialized on opposite sides of Saturn with orbital velocities corresponding to circular orbits at their initial orbital radius.  Saturn was then given an initial velocity determined by the constraint that the total linear momentum of the three-body system was zero so that the system's center of mass would remain stationary throughout the simulation.

\begin{figure}[ht!]
\centering
\includegraphics[scale=0.7]{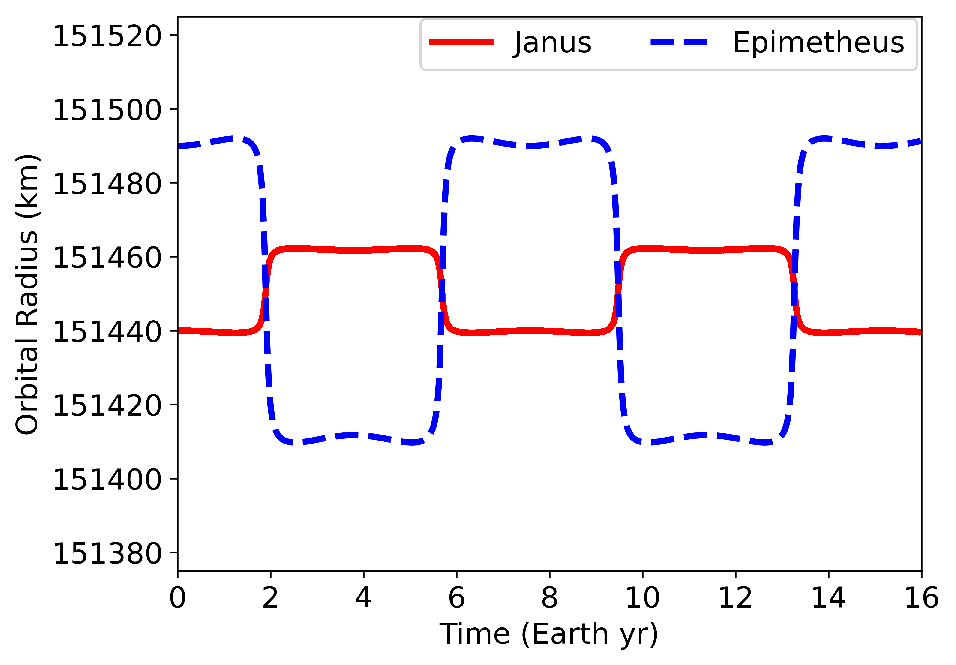}
\caption{Simulated orbital radii of Janus (solid line) and Epimetheus (dashed line) as a function of time with an initial orbital radius difference of 50 km.  The two moons exchange their relative ordering with respect to Saturn approximately every four Earth years, so that what had been the inner moon becomes the outer moon and vice-versa.}
\label{fig_radii_vs_t}
\end{figure}

The exchange is most easily visualized through graphs of the two moons' orbital radii as a function of time, as shown in Figure \ref{fig_radii_vs_t}.  In that image, one can see the orbital radii of the two moons exchanging order approximately every four Earth years, with the less massive moon migrating farther by a factor equal to the moon mass ratio.

\begin{figure}[ht!]
\centering
\includegraphics[width=15cm]{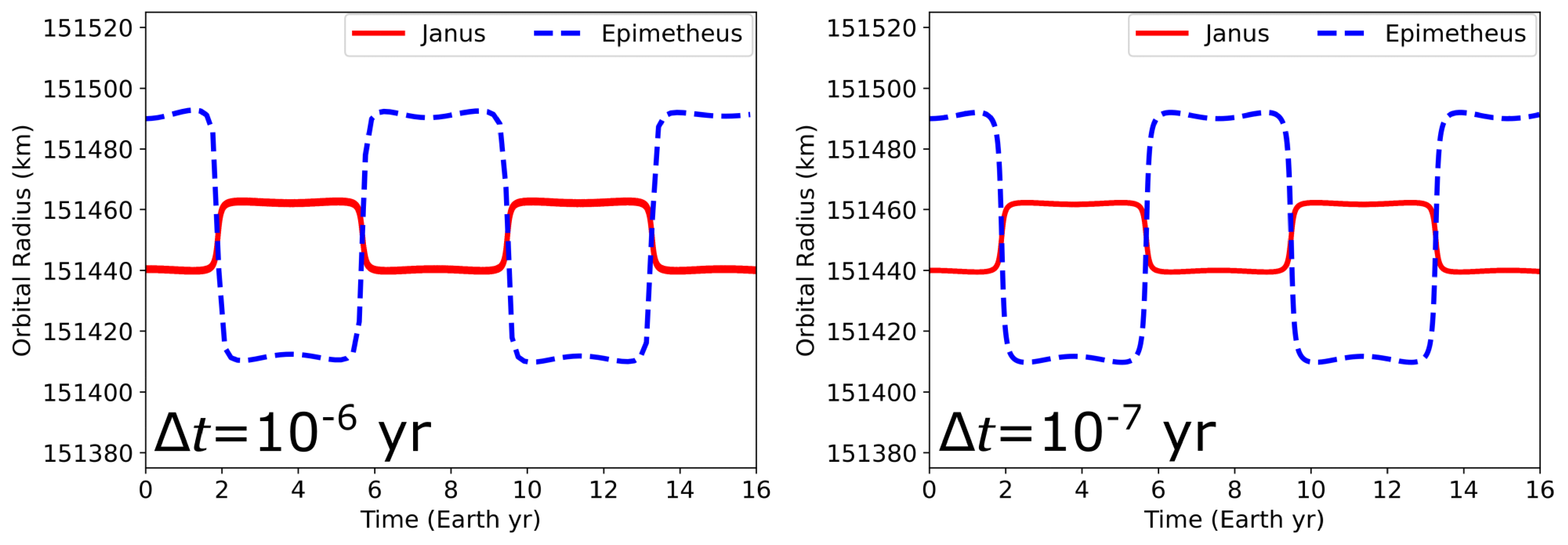}
\caption{A comparison between two $n$-body simulations of Janus and Epimetheus that are identical to one another except for the size of the integration timestep $\Delta t$.  The timestep in the left image is a factor of ten larger than in the right image, but the salient features of the two simulations are very similar, demonstrating that either timestep is sufficiently accurate for this particular set of simulation parameters.}
\label{fig_convergence}
\end{figure}

We also conducted a ``convergence test'' simulation featuring a smaller timestep $\Delta t$ to validate that the orbital solutions were not affected in important ways by the precise value of the timestep.  In Figure \ref{fig_convergence}, we compare two simulations that feature timesteps that differ by a factor of ten.  A very close examination reveals that the orbital radius lines in the leftmost (larger timestep) plot are slightly thicker than those of the rightmost (smaller timestep) plot, but the two simulations are remarkably similar, featuring orbital exchange periods and closest approach separations that match to four significant figures.  This comparison also allows us to estimate the uncertainties associated with parameters derived from our simulations.

\section{Results}
\label{results}

\begin{table}[h]
\caption{Analytical and Simulated Results, Observed Quantities for Co-Orbital Motion of Janus and Epimetheus}\label{table_results}%
\begin{tabular}{@{}llll@{}}
\toprule
  Quantity & Analytical  & Simulation & Approximate \\
  & Estimate  & Estimate & Observed Quantity \\
\midrule
Exchange Period (Earth years) & 3.8485 & 3.7905 $\pm$ 0.0002 & 4\footnotemark[1]\\
 Janus Post-Exchange Radius (km) & 151461.7 & 151461.7 $\pm$ 0.1 & 151460\footnotemark[2] \\
 Epimetheus Post-Exchange Radius (km) & 151411.7 & 151411.7 $\pm$ 0.1  & 151410\footnotemark[2]\\
Closest Approach (km) & 12530.1 & 12530.8 $\pm$ 1.3 & 15,000\footnotemark[1] \\
Duration of Exchange (Earth years) & (not estimated) & 0.35 - 0.45 & 0.3\footnotemark[3] \\
\botrule
\end{tabular}
\footnotetext[1]{NASA Science 2023}
\footnotetext[2]{Spitale et al. (2006) data from Cassini for years 2003 - 2005}
\footnotetext[3]{Lakdawalla (2006)}
\end{table}

In this section, we provide validation of our analytical solutions with $n$-body simulations.  Table \ref{table_results} compares our analytic predictions from Section \ref{analytical} to the numerical simulation that best reflects the real orbital parameters of Janus and Epimetheus by incorporating both the actual moon mass ratio (Janus is 3.6 times more massive than Epimetheus) and difference in initial orbital radii (50 km) (Jacobson et al. 2008).  For each quantity, the uncertainty values presented in the simulation estimates are derived from comparisons between convergence test simulations conducted using two different timesteps, as discussed in Section \ref{numerical}.  As expected, the largest mismatch between our analytical and simulation results occurs for the exchange period, which differs by approximately 1.5\% between the two approaches.  As discussed in Section \ref{analytical_period}, this is because our coarse analytical estimate did not include gravitational interactions between the two moons.  It is remarkable that the two numbers agree so well, given the assumptions of this simple model.

\begin{figure}[ht!]
\centering
\includegraphics[scale=0.7]{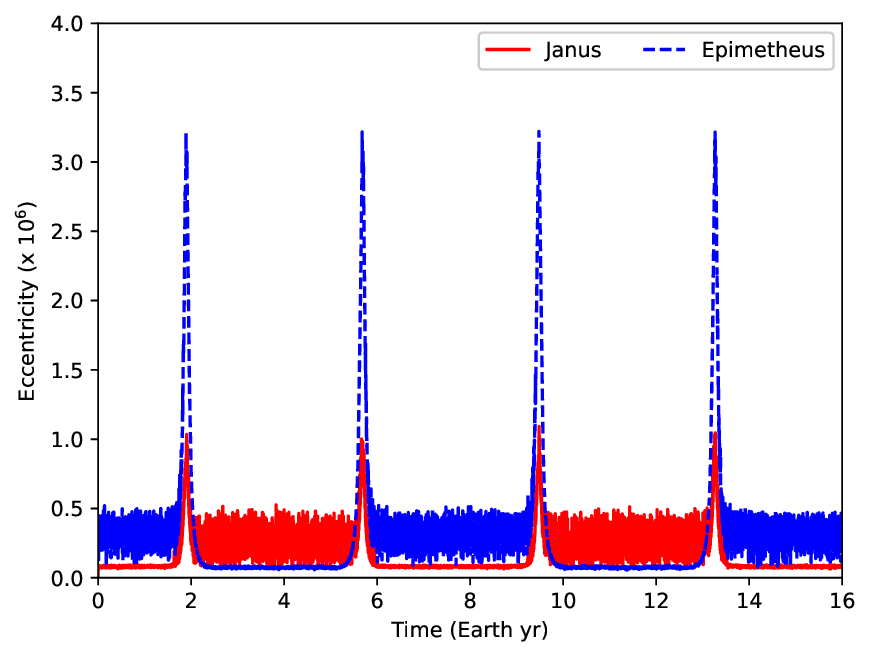}
\caption{The orbital eccentricities of Janus (lighter solid line) and Epimetheus (darker dashed line) as a function of time, as estimated from the periapsis and apoapsis distances for each object.  The time interval between adjacent spikes in eccentricity is representative of the co-orbital exchange period while the width of a given spike reflects the duration of that exchange.}
\label{fig_eccentricity}
\end{figure}

\begin{figure}[ht!]
\centering
\includegraphics[scale=0.7]{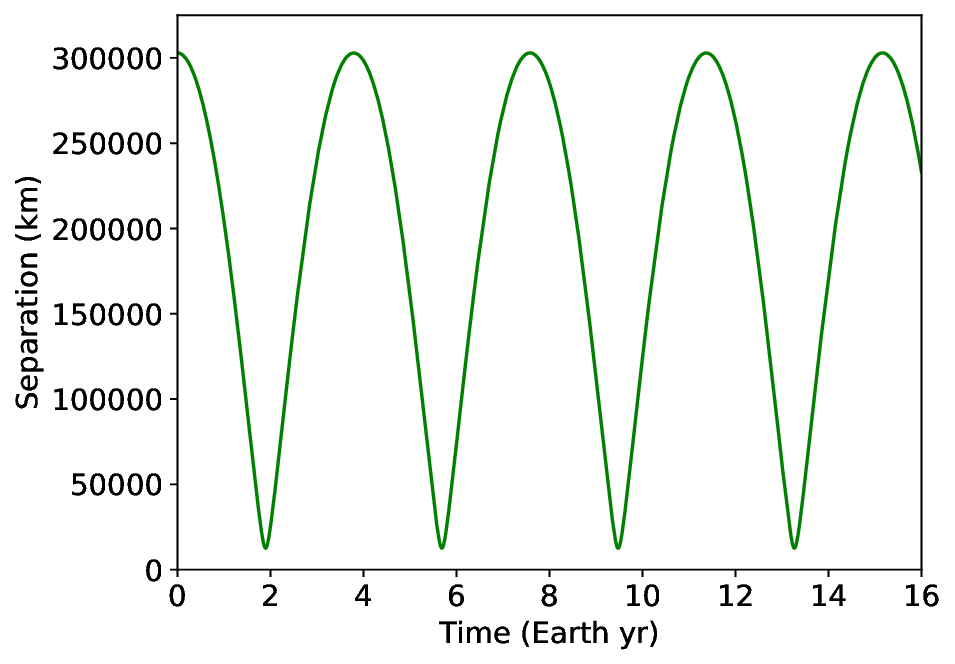}
\caption{The separation of Janus and Epimetheus as a function of time.  The time interval between adjacent separation minima (or maxima) is representative of the co-orbital exchange period.}
\label{fig_separation}
\end{figure}

Table \ref{table_results} also compares the analytical and simulated values of the post-exchange orbital radii and closest approach distances for the two moons.  For both quantities, the analytical predictions match the simulated values extremely well and fall easily within the relatively small uncertainties associated with the simulated values.  Table \ref{table_results} also provides an estimate for the duration of the orbital exchange, referring to the length of time over which the moons migrate noticeably in orbital radius.  This value, approximately 0.35 - 0.45 Earth years, can be estimated most directly from the moon eccentricities, measured in the simulations as
\begin{equation}e = \frac{r_a - r_p}{r_a+r_p}~,\end{equation}
where $r_a$ is the apoapsis distance (maximum distance between the moon and Saturn) and $r_p$ is the periapsis distance (minimum distance between the moon and Saturn) for a given orbit.  As seen in Figure \ref{fig_eccentricity}, the width of a single eccentricity spike represents the co-orbital exchange duration while the time interval between adjacent spikes is indicative of the exchange period.
A complementary view of the co-orbital exchange is provided in Figure \ref{fig_separation}, which shows the moon separation as a function of time.  In all cases, the simulated exchange period is consistently determined to be just under four Earth years for parameters corresponding to Janus and Epimetheus.

\begin{figure}[ht!]
\centering
\includegraphics[scale=0.7]{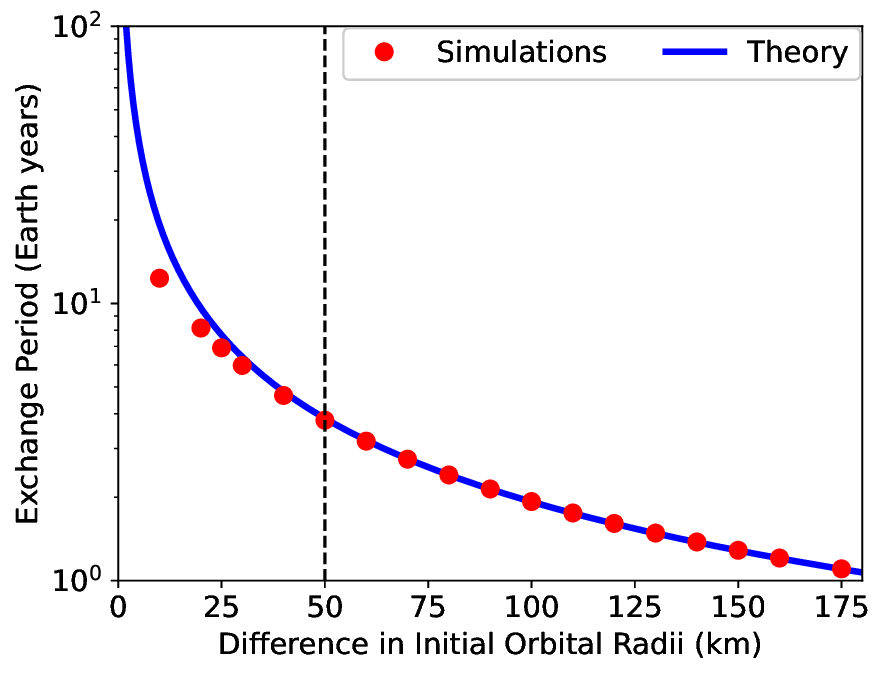}
\caption{Comparing the orbital exchange periods measured from simulations (dots) with those estimated using the analytic approach presented in Section \ref{analytical_period} (solid line) for the Janus-Epimetheus mass ratio (a specific case of the solutions presented in Figure \ref{fig_root_curves}).  The dashed vertical line represents the actual initial orbital radius difference for Janus and Epimetheus.  The simulations and analytic estimates match well for sufficiently large differences in initial orbital radius, but the coarse analytic prediction is less accurate for increasingly similar orbits on the left side of the graph.}
\label{fig_period}
\end{figure}

As discussed in Section \ref{numerical}, we also conducted an ensemble of simulations of co-orbital exchanges representing a variety of different values for the moons' initial orbital radii.  These simulations are particularly useful in testing our analytical estimates for hypothetical cases of co-orbital motion that do not precisely match the parameters of Janus and Epimetheus.  Figure \ref{fig_period} provides a comparison of the orbital exchange periods measured from simulations (dots) and those estimated using the analytic approach presented in Section \ref{analytical_period} (solid line).  The theoretical predictions and simulated results are in good agreement for sufficiently large values of the initial orbital radius difference, but there are clear differences present for orbital radius differences less than roughly 50 km.  As mentioned, these differences are attributable entirely to our analytic approach, which provided only a coarse estimate of the exchange timescale without actually incorporating the physics of exchanges.  Comparison with our simulation results illustrates that this approach matches the co-orbital exchange period of Janus and Epimetheus with a percent difference of 1.5\% and is a better fit for systems featuring larger initial orbital radius differences and worse for smaller differences in initial orbital radii.

\begin{figure}[ht!]
\centering
\includegraphics[scale=0.7]{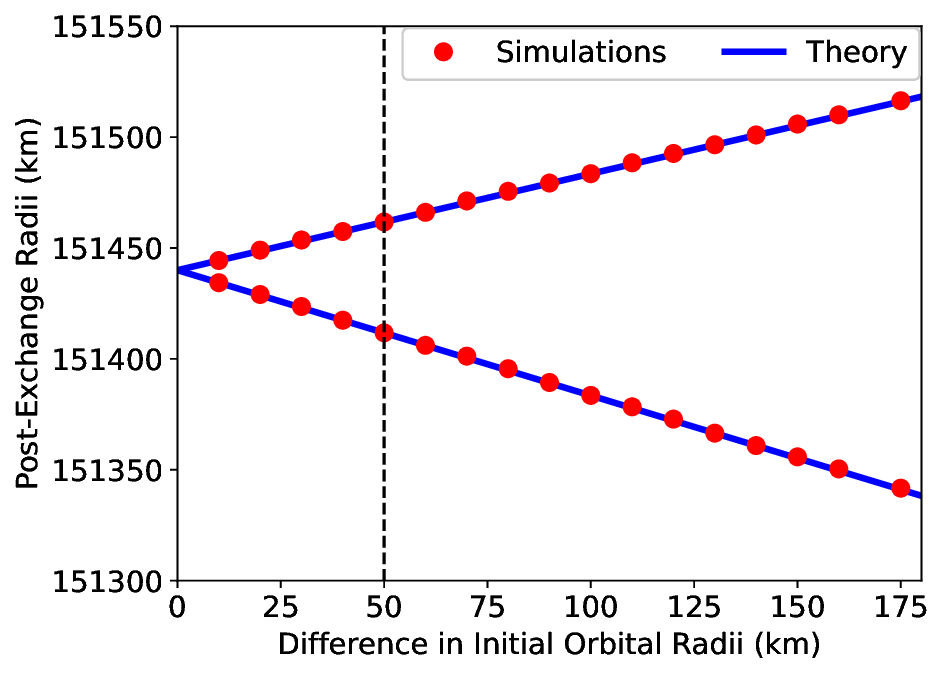}
\caption{Comparing the post-exchange orbital radii measured from our simulations (dots) with those derived from the analytic approach in Section \ref{analytical_radii} (solid lines) for the Janus-Epimetheus mass ratio.  The dashed vertical line represents the actual initial orbital radius difference for Janus and Epimetheus.  For all tested parameters, the theoretical predictions are visually indistinguishable from the simulated results.}
\label{fig_post_exchange_radii}
\end{figure}

In Figure \ref{fig_post_exchange_radii}, we compare the post-exchange orbital radii measured from our simulations (dots) with those derived from the analytic approach in Section \ref{analytical_radii} (solid lines).  Here, we see no perceptible differences between our predictions and the simulation results, and certainly any mismatches are contained well within the $\sim$ 0.1 km numerical uncertainty associated with the simulations.

\begin{figure}[ht!]
\centering
\includegraphics[scale=0.7]{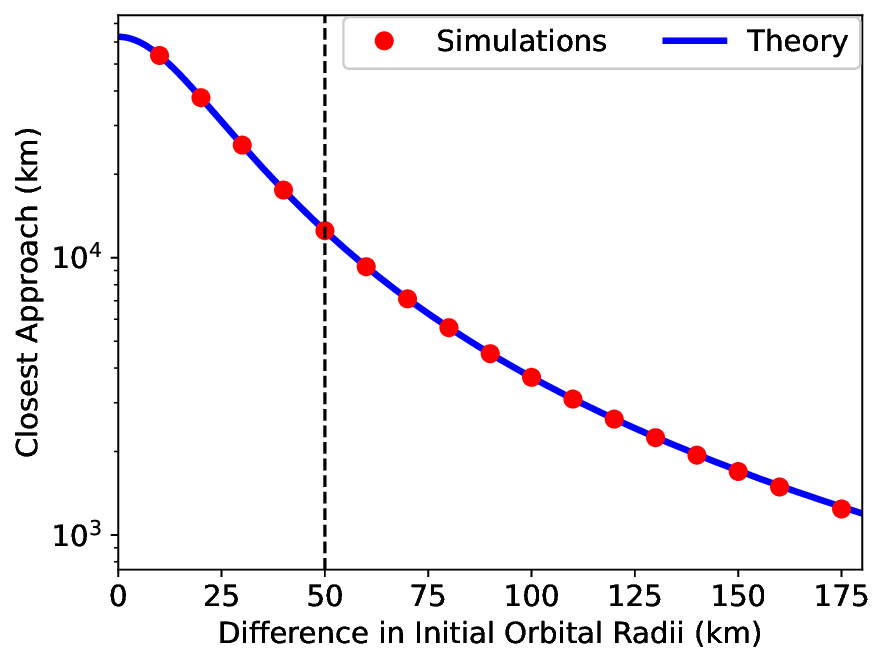}
\caption{Comparing the moon separations at closest approach measured from our simulations (dots) and those from the analytic approach discussed in Section \ref{analytical} (solid line) for the Janus-Epimetheus mass ratio.  The dashed vertical line represents the actual initial orbital radius difference for Janus and Epimetheus.  For all tested parameters, the theoretical predictions are visually indistinguishable from the simulated results.}
\label{fig_closest_approach}
\end{figure}

Figure \ref{fig_closest_approach} provides a comparison between the moon separation at closest approach measured from our simulations (dots) and that from the analytic approach discussed in Section \ref{analytical_closest_approach} (solid line).  Again, there is no perceptible difference between the analytical results and the simulations.  This is a particularly encouraging justification of the assumptions made in our analytic model, namely that the two moons actually must be moving with approximately the same speed at approximately the same distance from Saturn when they experience their closest approach.

Figure \ref{fig_closest_approach} also illustrates one particularly counter-intuitive aspect of co-orbital motion, namely that the two moons experience a closer approach for a larger initial difference in their orbital radii.  In other words, co-orbital moons that start off with more dissimilar orbits come closer to one another than those with more similar orbits.  If we extrapolate this trend to larger initial differences in orbital radii, we can estimate the parameters that would result in a closest approach equal to the sum of the two moons' physical radii, resulting in a collision.  For moons with the same mass ratio as Janus and Epimetheus, this collision would occur around an initial orbital radius difference of 500 km, which is approximately ten times larger than the difference in their actual current orbits.  This is an especially useful analytical result since an accurate simulation of such a collision would require a significantly smaller computational timestep and/or a higher-order code to capture the strong accelerations immediately preceding a collision.

\begin{figure}[ht!]
\centering
\includegraphics[width=12cm]{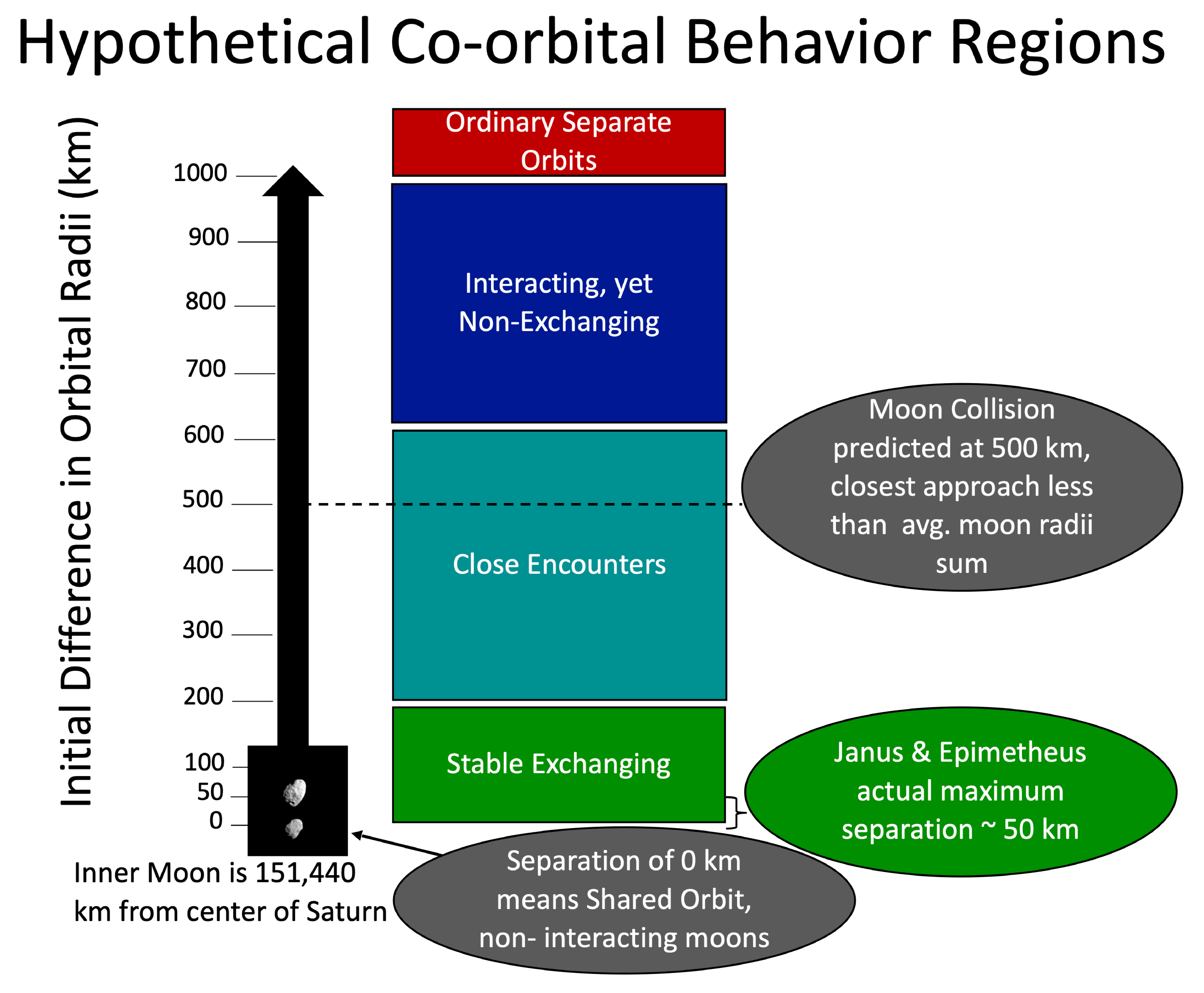}
\caption{A map of potential co-orbital moon behaviors for moons with the Janus-Epimetheus mass ratio.  Moons with similar orbital radii (bottom) experience stable exchanges, while those with more dissimilar orbital radii between 200 - 600 km are more likely to experience close encounters, collisions, or ejections.  Beyond 600 km, we expect that two moons could interact without exchanging until the point that they are so far apart that they would orbit independently.  Moon image adapted from Cassini-Huygens data (credit: NASA/JPL/Space Science Institute).}
\label{fig_regions}
\end{figure}

Generalizing from our analytic and simulation results, we find it useful to organize the parameter space of co-orbital motion into the regions depicted in Figure \ref{fig_regions}.  Starting at the bottom of the diagram, zero difference in initial orbital radius corresponds to two moons that trace out the same circular orbit without ever exchanging.  Just above that, the region marked ``Stable Exchanging'' contains all simulations that we have presented here, including the fiducial parameters corresponding to Janus and Epimetheus.
In the absence of an external perturbation or non-ideal effects such as tidal dissipation, such systems are expected to be stable over long timescales.  As we move up on the chart, we experience ``Close Encounters'' between the moons, which could potentially lead to a collision or an ejection of one of the two moons, depending on the exact parameters.  When the two moons have sufficiently different orbits, we reach the regime in which they are expected to be ``Interacting, yet Non-Exchanging,'' where they would perturb one another without exchanging their relative positions with respect to the planet.  Finally, at the top of the graph, there are moons that are always so far from one another that their orbits are essentially independent of one another.  Although the exact locations of the region boundaries are potentially adjustable (and would certainly vary with the exact moon mass ratio and inclusion of orbital resonances), it is conceptually useful to delineate the various regimes of behavior that can result from mutually interacting moons.

\section{Conclusions} 
\label{conclusions}
Our primary focus in this work has been to identify which co-orbiting moon parameters can be computed through the results of simple analytical arguments and which require numerical simulation.  Our analytical solutions require a small fraction of the time needed by full $n$-body simulations.  As such, these analytical solutions represent a valuable shortcut for determining orbital parameters such as orbital exchange period, post-exchange orbital radii, and separation at closest approach.  That said, $n$-body simulations do provide details of trajectories that cannot be found analytically.

As discussed in Section \ref{results}, our theoretical estimates of post-exchange orbital radii and moon separations at closest approach match the simulated values to a remarkably high degree of accuracy.  This is very useful since utilizing the closed-form expressions in Equations \ref{eq_post_exchange_radius1} and \ref{eq_post_exchange_radius2} for the post-exchange radii or root-finding Equations \ref{eq_closest_approach1} and \ref{eq_closest_approach2} for the closest approach would take at most a few minutes, compared to having to conduct full $n$-body simulations that would run for several hours to days at present.  If one were only interested in the post-exchange orbital radii or the moon separation at closest approach, it is clearly much easier to calculate these values analytically.

In the case of the exchange period, the analytical approach again gets a result much more quickly, although in this case with a potential tradeoff in accuracy.  Specifically, for parameters corresponding to Janus and Epimetheus, it is possible to predict the value of the exchange period to within a few percent without requiring a simulation.  For larger initial differences in orbital radius, the accuracy is even better.  For smaller initial differences in orbital radius, however, simulations would be required to deliver more accurate estimates of the exchange period.  Naturally, simulations would also be needed if one were interested in complete, accurate moon trajectories for co-orbital exchanges, which our analytic approach does not attempt to provide.

Our work also highlights some particularly counter-intuitive aspects of co-orbital motion, the most interesting of which is represented in Figure \ref{fig_closest_approach}, where one can clearly see that two co-orbiting moons experience closer approaches to one another for larger differences in initial orbital radii.  This provides a framework for predictions of close encounters or even collisions between co-orbiting moons, as discussed in Section \ref{results} and represented in Figure \ref{fig_regions}.

While the traditional ``horseshoe orbit'' representation presented in Figure \ref{fig_horseshoe} is useful as a means of subtracting out most of the moons' orbital motions, we found that the most effective visualizations of the exchanges come from time series of quantities such as orbital radius (Figure \ref{fig_radii_vs_t}), orbital eccentricity (Figure \ref{fig_eccentricity}), or moon separation (Figure \ref{fig_separation}).
A quick glance at any one of these figures can provide an approximate estimate of the orbital exchange period, while the orbital radius and eccentricity graphs also convey the 0.35-0.45 Earth year timescales over which most of the radial motion takes place in the exchange.
These plots serve as a reminder that the diverse timescales in the motion of Janus and Epimetheus include orbits shorter than one Earth day, relatively rapid migrations that take place over hundreds of Earth days, and an overall exchange period of roughly four Earth years.

While our simulations accurately reproduced the co-orbital exchanges of Janus and Epimetheus and explored some of the parameter space for similar systems with different initial orbital radius differences, there are still multiple promising avenues of related future work.
One avenue that we intend to explore is the inclusion of a realistic external perturber, which in this system would most likely correspond to the massive moon Titan.  Clearly the real system experiences exchanges even in the presence of Titan, but we seek to determine just how much of an influence Titan has on the simulated eccentricities of Janus and Epimetheus, which we found in our simplified three-body simulations to be much smaller than the $\sim 10^{-3} - 10^{-2}$ observed eccentricity values (Jacobson et al. 2008).

Another promising extension of this work is the exploration of mass ratios other than those for Janus, Epimetheus, and Saturn.  While the analytic work presented here should apply to any reasonable mass ratio corresponding to two low-mass moons orbiting a more massive planet, we have not fully explored how varying the mass ratio would adjust our expectations in Figure \ref{fig_separation}, for example, nor have we confirmed the validity of our analytical work across a wide range of mass ratios.
We are also interested in investigating more generally the relative importance of mass ratio, initial orbital radius difference, and proximity to the planet to co-orbital motion.
Such explorations will surely uncover aspects of co-orbital motion that, while perhaps not realized in our own solar system, could apply to undiscovered moons in our solar system or even those orbiting extrasolar planets.

\vspace{0.5 cm}
References
\vspace{0.25 cm}

Aarseth, S.J.: Gravitational $N$-Body Simulations. Cambridge University Press, Cambridge (2003). \\
\vspace{0.25 cm}

Aksnes, K.: The tiny satellites of Jupiter and Saturn and their interactions with the rings, in Stability of the Solar System and Its Minor Natural and Artificial Bodies. In: V. G. Szebehely (ed) Stability of the Solar System and Its Minor Natural and Artificial Bodies, p. 3–16. Kluwer, Boston (1985). \\
\vspace{0.25 cm}

Cooper, N. J., Renner, S., Murray, C. D., Evans, M. W.: Saturn's Inner Satellites: Orbits, Masses, and the Chaotic Motion of Atlas from New Cassini Imaging Observations. Astron. J. 419, 149-166 (2015).\\
\vspace{0.25 cm}

Cors, J. M., and Hall, G. M.: Coorbital periodic orbits in the three body problem. SIAM J. Appl. Dyn. Syst. 2, 219-237 (2003).\\
\vspace{0.25 cm}

\'Cuk, M., Hamilton, D. P., Holman, M. J.: Long-term stability of horseshoe orbits. Mon. Not. R. Astron. Soc. 426, 3051–3056 (2012).\\
\vspace{0.25 cm}

Dehnen, W., Read, J. I.: $N$-body simulations of gravitational dynamics. European Phys. J. Plus. 126, 1-28 (2011).\\
\vspace{0.25 cm}

Dermott, S. F., Murray, C. D.: The dynamics of tadpole and horseshoe orbits. Icarus 48, 12-22 (1981).\\
\vspace{0.25 cm}

El Moutamid, M., Nicholson, P. D., French, R. G., Tiscareno, M. S., Murray, C. D., Evans, M. W., French, C. M., Hedman, M. M., Burns, J. A.: How Janus' orbital swap affects the edge of Saturn's A ring?. Icarus 279, 125-140 (2016).\\
\vspace{0.25 cm}

Jacobson, R. A., Spitale, J., Porco, C. C., Beurle, K., Cooper, N. J., Evans, M. W., Murray, C. D.: Revised Orbits of Saturn's Small Inner Satellites. Astron. J., 135, 261-263 (2008).\\
\vspace{0.25 cm}

Kohlhase, C., Peterson, C. E.: The Cassini Mission to Saturn and Titan. ESA Bulletin 92, 54-68 (1997).\\
\vspace{0.25 cm}

Lakdawalla, E.: The Orbital Dance of Epimetheus and Janus. The Planetary Society. https://www.planetary.org/articles/janus-epimetheus-swap (2006). Accessed 04 Nov. 2023\\
\vspace{0.25 cm}

Llibre, J., Oll\'e, M.: The motion of Saturn coorbital satellites in the restricted three-body problem. Astron. Astrophys. 378, 1087–1099 (2001).\\
\vspace{0.25 cm}

Murray, C. D., Dermott, S. F.: Solar System Dynamics. Cambridge University Press, Cambridge (1999).\\
\vspace{0.25 cm}

NASA Science: Epimetheus. Saturn's Moons. https://science.nasa.gov/saturn/moons/epimetheus/ (page last updated October 2023). Accessed 04 Nov. 2023\\
\vspace{0.25 cm}

Neiderman, L., Pousse, A., Robutel, P.: On the co-orbital motion in the three-body-problem: existence of quasi-periodic horseshoe-shaped orbits. Commun. Math. Phys. 377, 551–612 (2020).\\
\vspace{0.25 cm}

Spitale, J. N., Jacobson, R. A., Porco, C. C., Owen, W. M., Jr.: The Orbits of Saturn's Small Satellites Derived from Combined Historic and Cassini Imaging Observations. Astron. J., 132, 692-710 (2006).\\
\vspace{0.25 cm}

Thomas, P. C.: Sizes, shapes, and derived properties of the saturnian satellites after the Cassini nominal mission. Icarus, 208, 395-401 (2010).\\
\vspace{0.25 cm}

Treffenst\"adt, L. L., Mour\~ao, D. C., Winter, O. C.: Formation of the Janus-Epimetheus system through collisions. Astron. Astrophys. 583, A80 (2015). \\
\vspace{0.25 cm}

Vanderbei, R. J.: Horsing around on Saturn. Ann. N. Y. Acad. Sci. 1065, 336-345 (2005).\\
\vspace{0.25 cm}

Yoder. C. F., Colombo, G., Synnott, S. P., Yoder, K. A.: Theory of motion of Saturn's coorbiting satellites,'' Icarus 53, 431-443 (1983).\\
\vspace{0.25 cm}

\end{document}